\documentclass [floatfix, showpacs, aps, pra, twocolumn]{revtex4-1}
\usepackage{amsmath}
\usepackage{graphicx}
\usepackage{bbm}
\usepackage{hyperref}
\usepackage{amssymb}
\usepackage[english]{babel}
\usepackage[latin1]{inputenc}
\usepackage{lmodern}
\usepackage[T1]{fontenc}

\graphicspath{{Figs/}}

\DeclareMathOperator{\e}{{\displaystyle e}}

\DeclareMathOperator{\vs}{{\displaystyle \mathbf{S}}}
\DeclareMathOperator{\vi}{{\displaystyle \mathbf{I}}}
\DeclareMathOperator{\si}{{\displaystyle \mathbf{S} \cdot \mathbf{I} }}

\graphicspath{{Figs/}}

\usepackage{amssymb}

\begin{document}

\title{Microtesla NMR J-coupling spectroscopy with an unshielded atomic magnetometer}

\author{Giuseppe Bevilacqua}
\address{DIISM, University of Siena - Italy}

\author{Valerio Biancalana}
\address{DIISM, University of Siena - Italy}

\author{Andrei Ben-Amar Baranga}
\address{Dep.\ of Electr.\ and Computer Engin., Ben-Gurion University of the Negev - Israel}

\author{Yordanka Dancheva}
\address{DSFTA, University of Siena - Italy}

\author{Claudio Rossi}
\address{DBCF, University of Siena - Italy}

\begin{abstract}
  We present  experimental data and theoretical  interpretation of NMR
  spectra of  remotely magnetized  samples, detected in  an unshielded
  environment  by means  of  a differential  atomic magnetometer.  The
  measurements are performed in an ultra-low-field at an intermediate regime,
  where  the  J-coupling  and  the  Zeeman  energies  have  comparable
  values and  produce rather  complex line sets,  which are satisfactorily interpreted.

\end{abstract}






\maketitle

\section{Introduction}
\label{Introduction}
Nuclear  magnetic  resonance   signals  are  typically  detected  with
inductive rf pickup  coils, and in conventional NMR  setups  signals
improve  approximately quadratically with  the magnetic field  B, because  both the
magnetization level and the Larmor frequency increase linearly with B.

In spectroscopic NMR applications the stronger  the field 
the better is the resolution of chemical shifts. However, the  spectral resolution
commonly  increases sub-linearly  with  B, because  the generation  of
stronger  fields  is  accompanied  by  larger  field  inhomogeneities,
causing greater instrumental broadening.

Conversely,  in NMR  spectroscopy   requiring the determination of intrinsic
splittings, rather than chemical shift measurements, weaker and  more homogeneous fields can be used.
In  fact, avoiding  intense  magnetic fields  helps  to improve  the
instrumental  resolution, although  it also rapidly leads  to poor  signal-to-noise 
ratio due to the above mentioned quadratic dependence.

This negative trend  can be counteracted by using
magnetometers  instead of  inductive pickup  coils, as  the former  make the
signal   strength   proportional   to   the   premagnetization
field alone. Additionally, when a  remote detection method is applied,  a strong  premagnetization field (whose  homogeneity is
unessential) can still be used, along with a low, homogeneous field  at the detection
stage. Non-inductive  detectors open the way to  unconventional NMR in
conditions  where the  nuclear  precession occurs  at arbitrarily  low
frequencies,  in  regimes  commonly known  as  zero to ultra-low-field
(ZULF-NMR) \cite{Ledbetter_2013}. ZULF-NMR  includes regimes  where the nuclear  spin coupling
becomes the  dominant (or  the only) term  in the Hamiltonian.

The state-of-the-art magnetic sensors is represented by superconducting quantum interference
devices   (SQUIDs)   and   atomic   magnetometers,   including   the latter's
spin-exchange-relaxation-free (SERF) implementation, the sensitivity of which
makes   them    competitive   with      SQUIDs,   surpassing   the
fT/$\sqrt{\mbox{Hz}}$ level \cite{sheng_13}.

Both kinds of detectors are used in the ZULF regime
\cite{weitekamp_prl_83, bernarding_jacs_06, savukov_prl_05,biancalana_arnmrs_13}, which has recently
enjoyed  renewed interest, partly due to its potential
application   in      imaging  \cite{savukov_jmr_13,   xu_pnas_06,
  mcdermott_jltp_04}. The use of 
optical  atomic  magnetometry  in  ZULF-NMR  detection  constitutes  an
interesting choice  due to its simplified setup, as, unlike SQUID, no cryogenics are
needed.  Moreover  atomic magnetometers can  operate  in
unshielded  environments,  which is  one of the  peculiarities  of the  work
presented in this paper. In this experiment, the magnetometer and the measured samples are merged in a magnetic field obtained by partially compensating the environmental one,  neither high permittivity nor thick conducting layers are used to passively shield continuous and low frequency magnetic field components.

From a spectroscopic viewpoint, the operation in zero magnetic field
\cite{weitekamp_prl_83,    ledbetter_pnas_08,    butler_jcp_13}    and
near zero magnetic field   \cite{ledbetter_prl_11}     conditions    leads   to
interpretation  schemes  \cite{appelt_cpl_07,theis_cpl_13}  that  are
complementary  to the conventional  ones:    the
dominant part of the  Hamiltonian describes the spin couplings, while the spin-field interactions are represented by perturbative terms.

Spectra  usually become     more  complex
 in an intermediate regime, where spin couplings and spin field interaction occur at comparable energies \cite{bernarding_jacs_06, appelt_pra_10}. In this case  the Hamiltonian shows  no perturbative terms
and, indeed, must be fully taken into account in modelling the  spectra recorded.

This   paper    presents the  experimental   results    and a  theoretical
interpretation  of  near zero field NMR  spectra  obtained  in  such an intermediate
regime,   recorded by means of a  differential optical magnetometer
operating  in an unshielded environment  that  detects  NMR signals  of
remotely  magnetized  samples at about $1$~T field \cite{belfi_josab_07, belfi_rsi_10}. 
The magnetometer operates at nearly $1~\mu$T field. Helmholtz coils 
and quadrupoles are used to control the magnetic field and to compensate its gradients, respectively.
In  particular,  we  report the spectra  of
protons in trimethyl-phosphate, whose  interaction with the P nucleus
causes a $^3J[H,P]$    splitting   of about 10 Hz \cite{dingley_cmr_01}, an
amount  comparable   with  their  Larmor   frequency  \cite{liao_sst_09,
  liao_rsi_10}.  Complex spectral  structures are recorded, whose main
features are  in good agreement with theoretical  evaluations based on
an analytic, non-perturbative analysis of the interaction Hamiltonian.

\section{Magnetometer experimental set-up}

\label{M setup}

\begin{figure}[htbp]
   \centering
  \includegraphics [width=8cm] {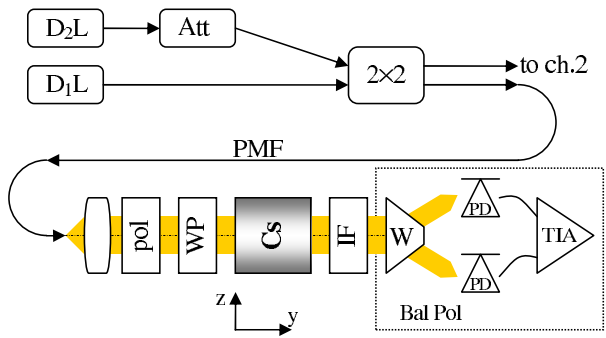}
  \caption{Schematics of one channel of  the magnetometer - D$_1$L  pump laser,
	D$_2$L  probe laser.  The magnetometer sensor (Cs cell)  is  illuminated  by the
	two fiber coupled (polarization maintaining fibres PMF)  radiations  mixed (at a ratio of $50\%$) in  the
    $2\times2$  coupler.  Before being mixed, the probe laser is attenuated down to $\mu$W level, while the pump is left at the mW level. After the Cs  cell, the pump radiation is stopped by means of  an  interferential  filter  (IF), and a balanced  polarimeter (BalPol), made of a Wollaston prism (W) and two photo-diodes (PD), measures the rotation of the probe beam polarization plane. The sample is positioned in the proximity of the sensor as depicted in Fig.~\ref{fig:ptm set-up}}
  \label{oursetup}
\end{figure}

The magnetometer has two identical arms for differential measurements.
Each  arm   (see
Fig.~\ref{oursetup}) contains  an illuminator providing  lights resonant
both with the $D_1$ line of  the Cs atoms (used to optically pump the
atoms) and with the $D_2$  line (used to  probe the atomic precession). The  two  radiations  are
collimated  into  beams  of 10~mm in diameter oriented along the $y$ direction, and a specifically 
 designed  multi-order wave-plate (WP) makes the pump radiation
circularly polarized, while leaving the probe one linearly polarized.

Each beam crossing the vapor is monitored by a balanced polarimeter.
The Cs cells  contain
$23~$torr  of N$_2$  as a  buffer gas. The Cs density is increased by warming it up to about  45~$^\circ$C  using an  alternating
current heater supplied at $50$~kHz,  a frequency higher than the
atomic Larmor frequency.

After  interacting  with the  atomic  vapour,  the  pump radiation (about 1~mW of power) is
blocked  by   an  interference  filter  (IF),  while   the  probe  beam 
polarization is analyzed by a  balanced polarimeter (BalPol).  A trans-impedance  amplifier  (TIA) converts  the
photo-current into a voltage signal,  which is then digitized
by means of a 16 bit  data acquisition (DAQ) card. The  signal recorded
is regarded  as the  real part of  an analytic signal  whose imaginary
part is inferred by means of a numeric Hilbert transform. The total phase $\theta(t)=\omega_0 t + \varphi$ of 
that  analytic  signal (which  corresponds  to  the  Larmor precession angle)  is    
analysed with  a linear regression routine. The phase $\varphi$ is weakly dependent upon time ($\dot \varphi \ll \omega_0$)
and reproduces the Faraday  rotation associated with
the local time-dependent magnetic field experienced by the weak and linearly  
polarized  probe  beam.

The  pump  radiation  is   generated  by  a  single-mode pigtailed distributed
feedback  diode laser, whose optical frequency is controlled
via the junction current, and is periodically (synchronously with the Larmor precession) made resonant to the
transition      $|^2S_{1/2}, F_g=3  \rangle \rightarrow
|^2P_{1/2} \rangle$ of  Cs, at  894~nm. 

Since the early 60's the synchronous optical pumping has been applied to drive precessing spins \cite{bell_prl_61}. 
In the present work in order to achieve maximum sensitivity, both the $D_1L$  detuning and the  modulation signal parameters have to be optimized. Due to the dual role of this radiation (hyperfine pumping and synchronous Zeeman pumping), the sensitivity depends non-trivially on these features, and this makes the optimization procedure not straightforward.

The probe laser is  a single-mode Fabry-Perot, pigtailed diode, whose light is resonant with the 
D$_2$ line of Cs at 852~nm. Its radiation probes the precession state at a low rate, being 
$2$~GHz blue-detuned from the maximum of the triplet
 $|^2S_{1/2}, F_g=4  \rangle \rightarrow |^2P_{3/2} \rangle$.

The arms of the magnetometer have a base-line of $5.6~\mbox{cm}$ (making it sensitive to the field variation along the $z$ direction). 
The magnetometer resonance line has a linewidth of the order
of ~20~Hz (limited by several relaxation processes, dominated by the spin-exchange collisions) and ensures a sensitivity
of $100~\mbox{fT}/\sqrt{\mbox{Hz}}$. Operating in an unshielded environment and in a gradiometric configuration
the maximum sensitivity is reached at about 100~Hz, where the environmental magnetic noise is
weaker and thus also the gradient/differential one gets lower.  The increase in the noise  at lower 
frequencies sets a sensitivity limit of a few pT/$\sqrt{\mbox{Hz}}$ at 1~Hz.

The magnetometer  operates  in a bias  magnetic   field  ranging  from
$100~\mbox{nT}$ to $6~\mu\mbox{T}$, whose direction ($z$, transverse to the laser beams) is established by three
pairs  of square Helmholtz coils with sides of 1.8~m     (see
Refs.\cite{belfi_josab_07,  belfi_rsi_10}  for additional  details).
The magnetic field Jacobian,  $G_{ij}=\partial B_i / \partial x_j$, is
zeroed  using five  quadrupole  magnets. One of the channels is used 
as a reference for the phase locking loop (PLL) developed to
decrease the common noise in the direction of the bias field (see Ref.\cite{belfi_rsi_10}).
The PLL locking loop has a bandwidth of the order of 200~Hz.

\section{NMR spectroscopy experimental set-up}
\label{NMR setup}

\begin{figure}
\includegraphics [width=8cm] {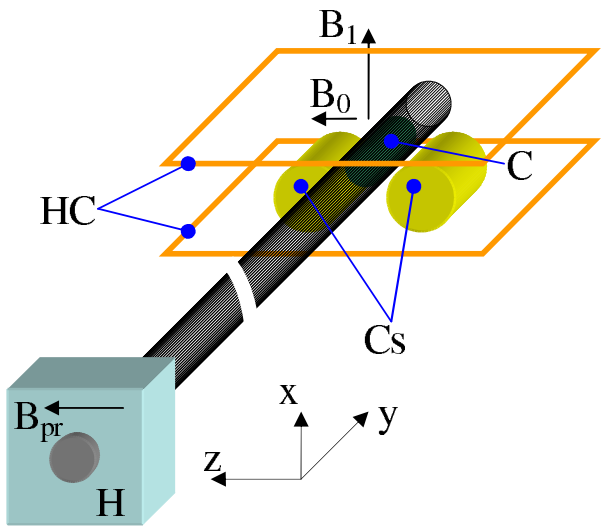}
\centering
\caption{Sample magnetization, shuttle  and positioning. The sample is
  contained in a polymer cartridge (C), which is inserted in a
  Halbach array  (H) and then pneumatically shuttled into
  the measurement region, 2~m away. Secondary 50~cm Helmholtz coils (HC) provide
  transverse $B_1$ pulses for nuclear spin manipulation.}
\label{fig:ptm set-up}
\end{figure}

The experimental  set-up for nuclear spin polarization, transport and
manipulation is represented in Fig.\ref{fig:ptm set-up}.
The  sample to  be  analyzed,  held in  a  5~cm$^3$ cartridge (3~cm in
length),  is
initially    placed   in  a  pre-polarizing   magnetic    field.   This
pre-polarization device is based on  Nd permanent magnets arranged in a
Halbach array, and provides a 1~T magnetic field in a cylindrical volume
of 25~mm in diameter and 50~mm in length. The sample is  pre-polarized for as long as about 5
times the  relaxation time measured. Subsequently it  is shuttled
into the proximity of the magnetometer  head. A detailed  description  of the  Halbach  assembly and  the pneumatic  sample-shuttle is  given  in Ref.\cite{biancalana_rsi_14}.  The travel time is of the order  of 130~msec.   The appropriate  position of the sample with respect to the sensor cells depends  on whether   the   longitudinal  relaxation   or   precession signal  need to  be
measured \cite{biancalana_jmr_09}.  For   the  results  presented  in  this   paper  the  sample
positioning is optimized for maximal precession signal measurement. 

\begin{figure}
\includegraphics [width=8cm] {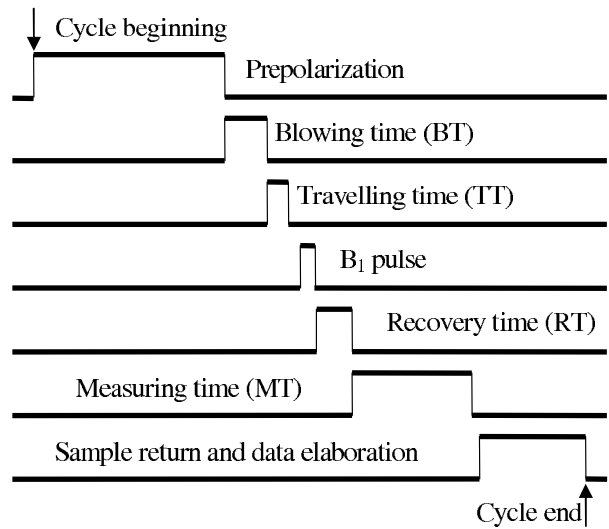}
\centering
\caption{Timing  of  the  cycle  for NMR  spectroscopy  (BT~=~$90$~ms;
  TT~=~$40$~ms; RT~=~$10-20$~ms; MT~=~$1-2$~s). The time interval
	in which the pneumatic shuttle pushes the sample towards the magnetometer
	head is called blowing time (BT).}
\label{timing}
\end{figure}

The measurements  are performed using an automated cycle procedure, whose timing is sketched in Fig.~\ref{timing}. 
The magnetic field along the sample displacement path is such that the
nuclear magnetization follows  it adiabatically. The exact positioning 
of  the  sample with  respect  to  the  dual channel  magnetometer 
is verified shot by shot, and  bad shots are automatically disregarded.

Operating  in a  low precession  field  makes it possible to manipulate nuclear spins by sudden field change. In
our experiment, a non adiabatic (square) $B_1$ pulse in the $x$ direction
is  used to tip  the nuclear  spins in  the $yz$  plane. 
The $B_1$ pulse is triggered   by  the   software   controlling  the shuttle  and   DAQ
operation, and is applied to a smaller Helmholtz pair (50~cm side), see
Fig.~\ref{fig:ptm  set-up}. This  produces a  transverse  magnetic field
exceeding the bias  field,  which  drives a  rotation  of the  sample
polarization at an angle adjustable via the pulse duration and amplitude. 
After being tipped with respect to the $z$ direction, nuclear spins precess around 
the same $B_0$ field where  the atomic spins of the sensors are also precessing.
The smaller gyromagnetic factor of the nuclei makes the nuclear signal
appear to be quasi static with respect to the atomic precession.

The  $B_1$ pulse  perturbs  the magnetometer, which then requires   $10-30$~ms  to recover its state of steady operation. Thus the pulse duration, summed with the recovery time and the sample transfer time, constitutes the total dead time after which  data acquisition may start. The NMR  signals presented  in this paper  are  average  signals over multiple   shots,   whose    superposition   requires   accurate   DAQ triggering with respect to the $B_1$ pulse.

The traces selected   are averaged  and analysed
with FFT-based evaluation of  their power spectral density.  The data
reported   in   this   paper   refer   to   water     and   to
trimethyl-phosphate (${\mbox[CH_3O]_3PO}$)  molecules, revealing in both cases only protons. This material is
chosen as a  test sample due to the presence of  two kinds of spin-$1/2$ nuclei, one of  which has a single  nucleus per molecule, which simplifies \cite{appelt_pra_10}    the    task     of   developing an analytical model.  

In the theoretical treatment, operating at intermediate field strengths renders several
assumptions  commonly  used in conventional NMR  spectroscopy inapplicable.  A  suitable model  is  derived here following  an
approach briefly  described in the  next section.

\section{Model}
\label{sec:model}
We assume that  the nine equivalent protons and  the phosphorus nucleus
can  be  represented  as   spin-$1/2$ nuclei  subjected  to J-coupling 
in an external magnetic field. 
To be more precise, we consider a Hamiltonian of the form ($\hbar =1$)
\begin{equation}
  \label{eq:H:form}
  H = - \gamma_S \, B\, S_z - \gamma_I \, B\, I_z + J \si ,
\end{equation}
where 
\begin{equation}
  \label{eq:def:S}
  \vs = \sum_{i=1}^9 \mathbf{s}_i
\end{equation}
is the total spin of the  protons, while $\vi$ is the spin-$1/2$ nuclei of
the phosphorus. The  axes are chosen in such a  way that the magnetic field
$B$ is in the $z$ direction. The $J$ is the exchange coupling constant 
($J=~^3J[H,P]$ in the experiment considered) and the $\gamma_i$ are the gyromagnetic constants. We also define  the
Larmor frequency  of the  two species as  $\omega_S = \gamma_S  B$ and
$\omega_I  = \gamma_I  B$.  The  Hamiltonian (\ref{eq:H:form})  can almost be
solved  analytically  \cite{bernarding_jacs_06, appelt_pra_10, abragam_61}.  In  fact there  are 3 
constants of motion which commute with  $H$, namely $F_z = S_z + I_z$,
$\vs^2$  and  $\vi^2$,   resulting  in  two  useful selection  rules
$\Delta F_z = 0$ and $\Delta S = 0$, while $\Delta I=0$ is trivial.

The simplest basis for the state vectors of the S-type spins can be chosen as 
\begin{equation}
  \label{eq:def:states}
  |\psi \rangle = | s_{1,z}\rangle \cdots | s_{9,z} \rangle
\qquad s_{i,z} = \pm 1/2
\end{equation}
however, only the  action of $S_z$ is readily written  in that basis. A
better choice is advisable, and, in fact, the basis states can be better 
labelled as
\begin{equation}
  \label{eq:label:altern}
  |\psi \rangle = | \eta; S, S_z \rangle, 
\end{equation}
where  $S(S+1)$, as  usual, is  the eigenvalue  of $\vs^2$,  $S_z$  is the
magnetic quantum  number, and $\eta$  is a label needed  to distinguish
states with the same $S$ and $S_z$. Upon inspection one finds that there
are  42  $S=1/2$, 48  $S=3/2$,  27 $S=5/2$,  8  $S=7/2$  and one
$S=9/2$ states, for a total of $512=2^9$. Including the $\vi$ part the states are $2\times 512 = 1024 = 2^{10}$ as can be noticed in the basis \eqref{eq:def:states}.    

The labelling of the global states is similar. Let us define
\begin{eqnarray}
  \label{eq:def:states:psi}
  | \psi_-(M; \eta S) \rangle & = & 
  |\eta S; S_z=M-1/2; I_z=1/2 \rangle \\
  | \psi_+(M; \eta S) \rangle & = & 
  | \eta S; S_z=M+1/2; I_z=-1/2 \rangle   .
\end{eqnarray}
These states are eigenstates of $\vs$, $S_z$ and $I_z$ and $F_z = | \psi_{\pm}(M; \eta S) \rangle = M | \psi_{\pm}(M; \eta S) \rangle $. 

The label
$\eta$ is not present in the Hamiltonian and thus the action of $H$ on
$|\psi_{\pm}(M; \eta S\rangle$ cannot depend on it \cite{abragam_61} and the matrix
elements are independent of $\eta$.  This is a consequence of the high
symmetry of the system, which leads to a Hamiltonian depending only on
the total spin operators. 

The action of $H$ on these states is straightforward 
\begin{eqnarray}
  \label{eq:action:H}
  H | \psi_-(M;\eta S) \rangle & = &
  a_- | \psi_-(M;\eta S) \rangle + 
  b | \psi_+(M;\eta S) \rangle \\
  H | \psi_+(M;\eta S) \rangle & = &
a_+
| \psi_+(M;\eta S) \rangle + 
b | \psi_-(M;\eta S) \rangle,
\end{eqnarray}
where 
\begin{equation}
  \label{eq:def:a:pm}
  a_{\mp }= -M\omega_S \pm ( M J + \omega_S - \omega_I)/2 -J/4,  
\end{equation}
and 
\begin{equation}
  \label{eq:def:b}
  b= \frac{J}{2} \sqrt{S(S+1) -M^2 -1/4}.  
\end{equation}

It follows  that we  have to  diagonalize many small  matrices not
larger than  a $2\times 2$  matrix.  Consider, for instance,  the states
with $S_z=S$ so that $M=S+1/2$, 
we have that $|\psi_-(S+1/2,S) \rangle$ is an eigenstate
because $|\psi_+(S+1/2,S) \rangle = 0 $ with energy $E_-^{e}(S) = -\omega_S S -
\omega_I/2   +J S/2$.    Also when $S_z= -S$ ($M= -S-1/2$),
$|\psi_+(-S-1/2, S)\rangle$ is an eigenstate because 
$|\psi_-(-S-1/2,S) \rangle = 0 $ with energy 
$E_+^{e}(-S-1/2,S) = \omega_S S +\omega_I/2 + J S/2$.

The other  ``non extremal'' states $\{  |\psi_-(M,S) \rangle, |\psi_+(M,S)
\rangle\} M=-(S-1/2),\ldots, S-1/2$ are coupled 
together and the energies read
$ E_{\pm }(S,M) = -(M \omega_S + J/4) \pm R/2$, where 
$ R = \sqrt{ ( \omega_S -  \omega_I + M J)^2 + J^2(S(S+1) -M^2 +1/4)}$.  
Using the  associated eigenstates and  standard methods, we calculate
the spectrum 
\begin{equation}
  \label{eq:spectro}
  \langle S_+(t) \rangle = \mathrm{Tr}( \rho(t) S_+ ) = 
  \mathrm{Tr}( \e^{-i H t} \rho(0) \e^{i H t} S_+)  
\end{equation}
as well as $ \langle I_+(t)  \rangle$, and thus the experimental one as
the real part of $\gamma_S \langle S_+(t) \rangle + \gamma_I \langle I_+(t) \rangle$.

The  initial density  matrix is  obtained  in two  steps, following  the
experimental procedure  outlined above. First the  sample is polarized
in a large field
\begin{equation}
  \label{eq:H:polar}
  H_{POLAR} = - \gamma_S B_P S_z - \gamma_I B_P I_z = - \bar{\omega}_S
  S_z - \bar{\omega}_I I_z,
\end{equation}
where $B_P  = 1$ T and it is  justifiable to neglect the  J coupling. After
some time in the polarization  region the density matrix of the sample
reaches the thermodynamic equilibrium value 
\begin{equation}
  \label{eq:rho:polar}
 \begin{split}
  \bar{\rho}(0)  & =  \frac{\e^{-\beta  H_{POLAR}}}{Z}  \approx  Z^{-1}(
  \mathbbm{1} - \beta\, H_{POLAR}) \\
  & = 
  \mathbbm{1} + \beta \bar{\omega}_S \,S_z + \beta \bar{\omega}_I \,I_z,
\end{split}
\end{equation}
where $\beta =1/(k_B T)$ is the inverse temperature and we drop the
normalization factor $Z$ because it is just a scale factor in the spectrum as
can be seen from \eqref{eq:spectro}. 

In the second step, which takes places in the measuring region, the spins are exposed for
a time $t$ to a ``rotation'' field $B_1$ as well as to the bias field
$B_0$.  Thus, neglecting the J-coupling effects, they sense the Hamiltonian
\begin{equation}
  \label{eq:H:rotaz}
  \begin{split}
  H_R & = - \gamma_S (B_1 S_x + B_0 S_z) - \gamma_I (B_1 I_x + B_0 I_z) \\
  & = - (\omega_{S,R} S_x + \omega_{S,0} S_z) 
  - (\omega_{I,R} I_x + \omega_{I,0} I_z) 
\end{split}
\end{equation}
and the final density matrix is 
\begin{equation}
  \label{eq:fin:rho}
  \rho(0) = \e^{-i H_R t} \bar{\rho}(0)\e^{i H_R t}. 
\end{equation}
Defining 
$\Omega_S = \sqrt{ \omega_{S,R}^2 + \omega_{S,0}^2}$, $\cos \chi_S = \omega_{S,R}/\Omega_S$
and $\sin \chi_S = \omega_{S,0}/\Omega_S$ the transformed $S_z$ can be
written as 
\begin{equation}
  \label{eq:fin:Sz}
  \begin{split}
    \e^{-i H_R t} S_z \e^{i H_R t} = & (1 - \cos (\Omega_S t)) \sin \chi_S
  \cos \chi_S \;S_x \\
  & + \cos \chi_S \sin (\Omega_S t) \; S_y \\
  & + (\cos(\Omega_S t) \cos^2 \chi_S + \sin^2 \chi_S) \; S_z .
\end{split}
\end{equation}
Using this expression and a similar one for $ \e^{-i H_R t} I_z \e^{i H_R t}$, the 
explicit form of $\rho(0)$ is recovered. 

\begin{figure}
\includegraphics [width=9cm] {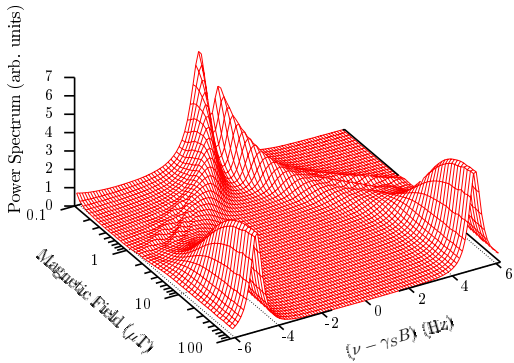} 
\centering
\caption{Simulated spectrum of trimethyl-phosphate  as a function of the magnetic field,  evaluated for $^3J[H,P]=63$~rad/s. For convenience, the frequency axis reports the detuning from $\gamma_s B$.
The curve is obtained by convolving the calculated stick spectrum (made of 100 components) with a Lorentian profile.  Going towards lower fields, the high-field doublet evolves in broader structures, which eventually start merging in a narrow structure at zero field.  Notice that the plot reproduces only the part of the spectrum showing the evolution of the hydrogen doublet. The small increase visible in the leftmost corner at vanishing field is due to other peaks.
\label{fig:3dgeneral}}
\end{figure}

The model reproduces the trimethyl-phosphate spectrum, which  is reported in Fig.~\ref{fig:3dgeneral} over a wide range of magnetic fields (3 decades, from the near-zero up to the geomagnetic level). In this figure the trivial linear slope of the frequency dependence on the magnetic field is removed, i.e. the frequency axis reports the detuning from the linear $\gamma_s B$ value, in order to better visualize the region of interest. The spectrum contains a large set of discrete peaks at the differences of many couples selected among 1024 eigenvalues, with possible degeneracies. The selection rules reduce the total amount of transitions to 100 \cite{appelt_pra_10}).
The smooth curves reported here are obtained by convolution of such discrete spectra with a Lorentzian profile.
Figure~\ref{fig:3dgeneral} shows the well-known high-field doublet-like spectrum at magnetic fields of about 100~$\mu$T and larger, with a field-independent separation of the two peaks (the
$^3J[H,P]/ h$ value). In the intermediate field range  (from tens of $\mu$T down to hundreds of nT) both hetero-nuclear spins and spin-field coupling determine a complex (and thus lower line intensity) spectrum with a field-dependent doublet. Here the peak separation is smaller at weaker fields. In this range, at the lower field intensity the complex spectral structures start to be resolved, rendering the doublet structure progressively less recognizable. Eventually, when approaching the near-zero-field regime, all the lines start to merge into a narrow structure, and another component starts appearing at near zero field. Only the tail of this second feature is visible in Fig.~\ref{fig:3dgeneral}, where it appears as a faint increase in the leftmost corner.

\section{NMR spectroscopy in trimethyl-phosphate}
\label{NMR spectroscopy}

\begin{figure}
\includegraphics [width=6cm] {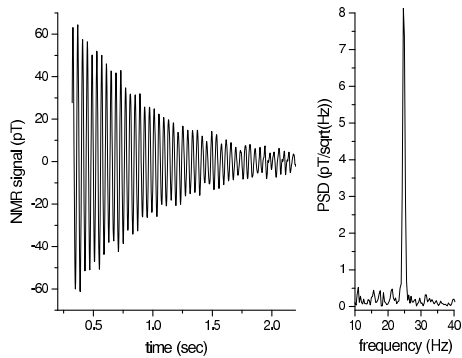} 
\centering
\caption{NMR signal signal from water protons. The signal is produced by 3.5~ml 
demineralized water premagnetized for 5~sec and measured at 0.58~$\mu$T. 
This plot is obtained by averaging over 20 cycles. First-order bandpass 
filtering with cut-off frequencies at 17 and 37~Hz is applied to better visualize the time-domain trace.}
\label{fig:fidwater}
\end{figure}

Preliminary  tests of the  whole experimental  setup are  performed by
acquiring proton  signal from water  samples. This step   also helps to
refine  the calibration  of the  $B_1$ coils  and to  adjust  the time
sequences. As  an example,  Fig.~\ref{fig:fidwater} shows a  precession signal  from a
3.5~ml  water sample both  in the
frequency  (as power  spectral density)  and in  the time  domain. The
time-domain  plot is band-pass filtered to reduce the low-frequency  noise and
other peaks of technical noises. Working in the nT-$\mu$T range the
magnetometer provides proton spectral NMR resolution better than 0.1~Hz due to magnetic
field homogeneity - better than B/$\Delta$B=$10^3$ (at $1$~$\mu$T bias field) over the sample volume of $6.3~cm^3$. 
The stated magnetic field homogeneity is obtained compensating the first order dc magnetic field gradients
with a simple and inexpensive shimming system made out of permanent magnets and large-area anti-Helmholtz coils. Thus
spin relaxation and diffusion processes can be characterized with high level of accuracy.

\begin{figure}
\includegraphics [width=8cm] {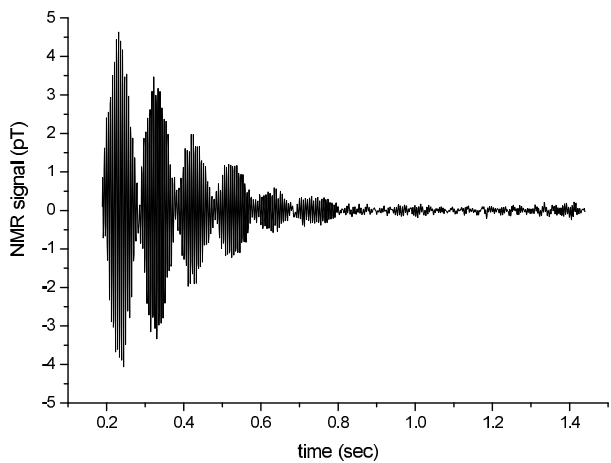}
\centering
\caption{NMR signal obtained with a 4.0~ml trimethyl-phosphate sample, 
in a bias magnetic field of 4.47~$\mu$T, with a prepolarization time of $3$~seconds. 
The signal is an average over 500 shots. The data are filtered with a first 
order band-pass filter with cut-off frequencies at 170 -- 220~Hz.}
\label{fig:fid192hz}
\end{figure}

The scalar coupling of the heteronuclear spin system as a function of the magnetic field
is measured in a sample of trimethyl-phosphate. One example of the precession signal obtained at 4.47~$\mu$T is shown in Fig.~\ref{fig:fid192hz}. 
At the highest  field strength investigated
(of the order of 5~$\mu$T), the proton Larmor frequency is of the order of 200~Hz, 
the high field approximation  holds, and the proton spectrum appears
as a  doublet. The precession signal can be approximated as a decaying beat (see Fig.~\ref{fig:fid192hz}) of two
spectral components, with a separation given by the $^3J[H,P]$  constant, which amounts to about 10~Hz
\cite{liao_sst_09,  liao_rsi_10, mcdermott_sc_02}.  

\begin{figure}
 \includegraphics [width=8cm] {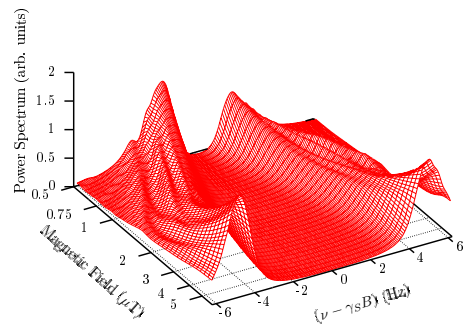}
 \centering
 \caption{Simulated spectrum of  trimethyl-phosphate in the intermediate field range. The same curve shown in Fig.~\ref{fig:3dgeneral}  is here plotted for the field interval investigated experimentally. This more detailed view shows that, when approaching the intermediate regime, the two high-field components become broader and start slowly approaching each other. At weaker fields they evolve in two weaker groups of partially resolved lines, which keep approaching faster and faster.}
 \label{fig:3dint}
\end{figure}

A simplified model of the signal made of only two components, makes it possible to fit the precession signal in Fig.~\ref{fig:fid192hz}, provided that two different decay times are considered. This simplified best-fit estimation gives time constant values of 190~ms and 240~ms for the high- and low-frequency components, respectively. This difference 
reflects the complexity of the theoretical spectrum presented in Fig.~\ref{fig:3dint}
where the right-hand component of the spectrum (corresponding to the high-frequency component of the experimental spectrum) shows a bigger linewidth. The theoretical curves presented in Figs.~\ref{fig:3dgeneral} and \ref{fig:3dint} are actually obtained by convolving the discrete spectrum produced by the model with a Lorentzian profile whose width is $\Gamma=1/500$~ms,  so that the width in excess is definitely set by the unresolved spectral structures.

Figure~\ref{fig:3dint} shows that, besides their apparent broadening due to the separation of the unresolved components that eventually appear as secondary peaks, at weaker fields the two components of the doublet evolve into two broader and weaker groups of lines, whose spectral separation progressively decreases.

\begin{figure}
 \includegraphics [width=8cm] {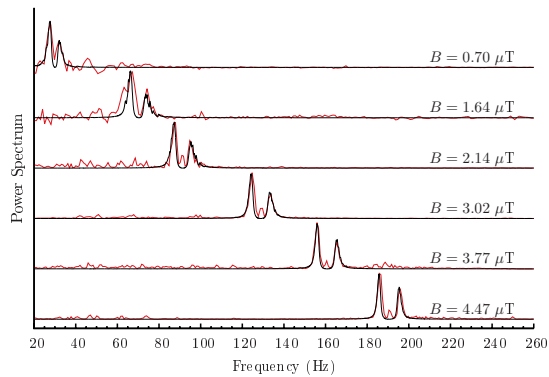}
 \centering
 \caption{Trimethyl-phosphate experimental and theoretical NMR spectra recorded at different bias magnetic fields. 
 Each trace is an average of about 500 shots. The average trace is only band-pass filtered
 (1-st order filter) around the doublet structure, with a cut-off frequency gap of 50~Hz.}
 \label{fig:comparison}
\end{figure}

Our aim is to prove that a completely unshielded optical magnetometer is a good tool for
high resolution NMR spectroscopy even in this intermediate-field regime, where the 
richness of the ${\mbox[CH_3O]_3PO}$ spectrum makes  the interpretation of  results difficult.  
Figure~\ref{fig:comparison} shows the experimentally measured
spectrum together with the theoretical one as a function of the bias field. 
The small spectral component between the peaks of the doublet structure is likely due to the presence
of water protons in the sample holder.
The height of the experimental peaks is normalized, which is why a decrease of the signal-to-noise is observed at lower magnetic field.
At a 700~nT bias field the overlapping of the two main structures
is  pronounced. A comparison between the experimental and theoretical curves is shown
in Fig.~\ref{fig:comparison}, calculated with $^3J[H,P]=10$~Hz.
There is excellent agreement between the experimental curves and the theoretical evaluation. It is worth noting that, apart from a phenomenological width of the profile used for the convolution of the discrete spectrum, the model includes only the $^3J[H,P]$ as a free parameter. Thus the close correspondence between theory and experiment demonstrates that useful information and accurate evaluations can also be inferred  from measurements performed in this intermediate regime.

\section{Conclusion}
In this paper we report ZULF-NMR spectroscopy results obtained
with a trimethyl-phosphate sample of  4~ml in volume premagnetized in a
1~T field.  The  experiment  is based  on  remote-detection of  precession signal in an unshielded environment by means of a
differential optical  magnetometer.  The system  is tested at intermediate  field strengths,
where the  $^3J[H,P]$ coupling  and the spin-field  interaction are  of the
same order  of magnitude. An analytical  model is  developed to  interpret  the 
spectra observed, obtaining consistent predictions. 

The work presented here shows the potential of  optical magnetometry in ZULF-NMR spectroscopy even  in an unshielded environment and when operating in a disadvantageous regime, where a complex spectral structure renders signal analysis a challenging task.

\section* {Acknowledgments}
The  authors acknowledge  the valuable  technical support  of Leonardo
Stiaccini and the financial support of Italian Ministry for Research
(FIRB  project  RBAP11ZJFA005). The authors thank  E.~Thorley of Language Box (Siena) for revising the English in the manuscript.

\end{document}